\documentclass[prd,aps,showpacs,tightenlines]{revtex4}  %%preprint,
\usepackage{mathrsfs}
\usepackage{amsmath}
\usepackage{amssymb}
\usepackage{epsfig}
\usepackage{graphicx}
\usepackage{booktabs}
\usepackage{multirow}
\usepackage{subfigure}
\usepackage{color}
\begin{document}
%%%%%%%%%%%%%%%%%%%%%%%%%%%%%%%%%%%%%%%%%%%%%
%\renewcommand{\arraystretch}{0.5}
\newcommand{\psl}{ p \hspace{-1.8truemm}/ }
\newcommand{\nsl}{ n \hspace{-2.2truemm}/ }
\newcommand{\vsl}{ v \hspace{-2.2truemm}/ }
\newcommand{\epsl}{\epsilon \hspace{-1.8truemm}/\,  }

%%%%%%%%%%%%%%%%%%%%%%%%%%%%%%%%%%%%%%%%%%%%%%%%%%%%%%

\title{Semileptonic decays of $B_c$ meson to S-wave charmonium states  in the perturbative QCD approach}
\author{Zhou Rui$^1$}\email{jindui1127@126.com}
\author{Hong Li$^1$}
\author{Guang-xin Wang$^1$}
\author{Ying Xiao$^2$}
\affiliation{$^1$College of Sciences, North China University of Science and Technology,
                 Tangshan 063009,  China\\
                 $^2$College of Information Engineering, North China University of Science and Technology,
                 Tangshan 063009,  China}
\date{\today}

\begin{abstract}
Inspired by the recent measurement of the ratio of $B_c$ branching fractions to $J/\psi \pi^+$ and $J/\psi \mu^+\nu_{\mu}$ final states at the LHCb detector, we study the semileptonic decays of $B_c$ meson to the S-wave ground and radially excited 2$S$ and 3$S$  charmonium  states with the perturbative QCD approach.
After evaluating the form factors for the transitions $B_c\rightarrow P,V$, where $P$ and $V$ denote
pseudoscalar and vector S-wave charmonia, respectively, we calculate the branching ratios for all these semileptonic decays. The theoretical uncertainty of hadronic input parameters are reduced by utilizing  the light-cone wave function for the $B_c$ meson. It is found that the predicted branching ratios range from $10^{-7}$ up to $10^{-2}$ and could be measured by the future LHCb experiment. Our prediction for the ratio of branching fractions $\frac{\mathcal {BR}(B_c^+\rightarrow J/\Psi \pi^+)}{\mathcal {BR}(B_c^+\rightarrow J/\Psi \mu^+\nu_{\mu})}$
is in good agreement with the data. For $B_c\rightarrow V l \nu_l$ decays,  the relative contributions of the longitudinal and transverse  polarization are
discussed in different momentum transfer squared regions.
These predictions will be tested on the ongoing and forthcoming experiments.
\end{abstract}

\pacs{13.25.Hw, 12.38.Bx, 14.40.Nd }

\keywords{ }

\maketitle

\section{Introduction}

Recently, the LHCb Collaboration has measured the semileptonic and hadronic decay rates of the $B_c$ meson and obtained $\frac{\mathcal {BR}(B_c^+\rightarrow J/\Psi \pi^+)}{\mathcal {BR}(B_c^+\rightarrow J/\Psi \mu^+\nu_{\mu})}=0.0469\pm 0.0028(stat)\pm 0.0046(syst)$ \cite{prd90032009}. It is a motivation to investigate the $B_c$ meson semileptonic decays to charmonium which are easier to identify in experiment. Indeed, both the CDF and the D0 Collaboration have measured the lifetime of the $B_c$ meson through its semileptonic decays \cite{prl812432,prl97012002,prl102092001}. More recently, the LHCb Collaboration gave a more precise measurement of its lifetime using semileptonic $B_c\rightarrow J/\psi \mu \nu_{\mu} X$ decays \cite{epjc742839}, where $X$ denotes any possible additional particles in the final states. At the quark level, the semileptonic decays of the $B_c$ meson driven by a $b\rightarrow c$ transitions, where the effects of the strong interaction can be separated from the effects of the weak interaction into a set of Lorentz-invariant form factors. It may provide us with the information as regards the Cabibbo-Kobayashi-Maskawa (CKM) matrix elements $V_{cb}$ and the weak $B_c$ to charmonia transition form factors.

There are many theoretical approaches to the calculation of $B_c$ meson semileptonic decays to charmonium. Some of them are: the nonrelativistic QCD \cite{prd87014009}, the Bethe-Salpeter relativistic quark model \cite{14113428}, the relativistic quark model  \cite{prd68094020,prd63074010}, the light-cone QCD sum rules approach \cite{prd77054003,epjc51833}, the covariant light-front model \cite{prd79054012}, the nonrelativistic
quark model \cite{prd74074008}, the QCD potential model \cite{prd61034012,11094468}, and  the light front quark model \cite{prd89017501}. The perturbative QCD (pQCD) \cite{prl744388} is one of the recently developed theoretical tools based on QCD to deal with the nonleptonic and semileptonic B decays. So far the semileptonic $B_{u,d,s,c}$ decays have been studied systematically in the pQCD approach \cite{prd86114025,prd87097501,prd89014030,csb59125}. One may refer to  the review paper \cite{csb593787} and the  references therein.

In our previous work \cite{prd90114030,epjc75293}, we analyzed the two-body nonleptonic decays of the $B_c$ meson with the final states involving one S-wave charmonium using the perturbative QCD based on $k_T$ factorization. By using the harmonic-osillator wave functions for the charmonium states, the obtained ratios of the branching fractions are consistent with the data and other studies.  Especially some of our
predictions were well tested by the recent experiments at ATLAS \cite{epjc764} and  LHCb \cite{prd92072007}, which may indicate that the harmonic-oscillator wave functions for S-wave charmonium  work well.

In this paper, we extend our previous pQCD analysis to the semileptonic $B_c$ decay such as $B_c\rightarrow (\eta_c(nS), \psi(nS))l \nu$ (here $l$ stands for the leptons $e$, $\mu$ and $\tau$) with the radial quantum number $n=1,2,3$, while the higher 4$S$ charmonia are not included here since their properties are still
not understood well.
The semileptonic decays $B_c\rightarrow (J/\psi,\eta_c)l \nu$  have
been studied in pQCD \cite{cpc37093102}, compared to which the new
ingredients of this paper are the following. 

(1) Instead of the traditional zero-point wave function for the $B_c$ meson, the   light-cone wave function which was well developed in Ref. \cite{prd89114019} is employed in order to reduce the uncertainties caused by the  hadronic parameters. In addition,  the charmonium distribution amplitudes are also extracted from the correspond Schr$\ddot{o}$dinger states for the harmonic-oscillator potential. (2) Here, the momentum of the spectator charm quark is proportional to the corresponding meson momentum. In Ref. \cite{cpc37093102}, the charm quark in the  $B_c$ meson carries a momentum with only the minus component. That is, its invariant mass vanishes, while  the charm quark in the final states is proportional to
the charmonium meson momentum and  its invariant mass does not
vanish.  This substantial revision will render our analysis more consistent. (3) We updated  some input hadronic parameters according to the Particle Data Group 2014 \cite{pdg}. (4) Besides including the $B_c\rightarrow (J/\psi, \eta_c) l \nu$
decays, the $B_c\rightarrow P/V(2S,3S) l \nu$ decays are also investigated, where it
is theoretically easier compared with that of nonleptonic decays. Our goal is to provide a ready reference to the
existing and forthcoming experiments to compare their data
with the predictions in the pQCD approach.

The paper is organized as follows. In Sect. \ref{sec:kw} we define kinematics and describe the wave functions of  the initial and final states,
while the analytic expressions for the transition form factors  and the  differential decay rate of the considered decay modes are given in Sect. \ref{sec:cul}. The numerical results and relevant discussions are given in Sect. \ref{sec:res}. The final section is theconclusion. The evaluation of the  3$S$ charmonium distribution amplitudes  is relegated to the appendix.
\begin{figure}[!htbh]
\begin{center}
\vspace{-2cm} \centerline{\epsfxsize=12 cm \epsffile{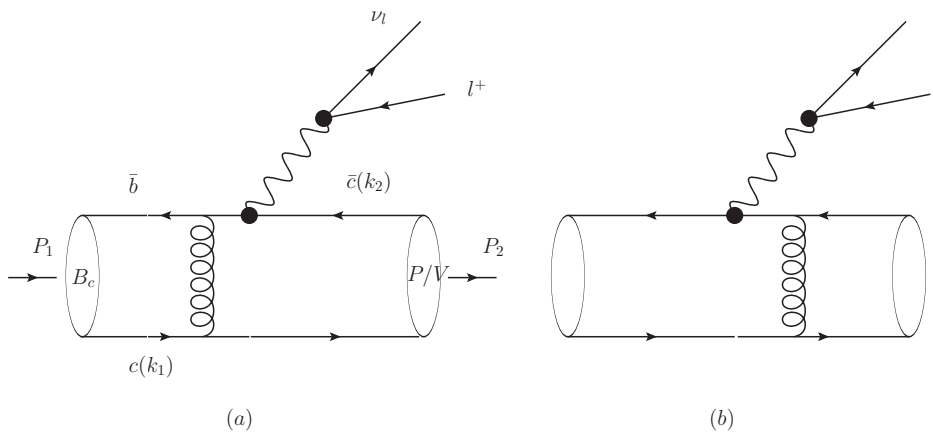}}
\vspace{-5.6cm} \caption{The leading-order Feynman diagrams for the
 semileptonic decays  $B^+_c\rightarrow P/V l^+ \nu_l$ with $l=(e,\mu,\tau)$.}
 %, where $X_{c\bar{c}}$ stand for the final state charmoniu meson}
 \label{fig:semi}
 \end{center}
\end{figure}
\section{ Kinematics and the wave functions}\label{sec:kw}

It is convenient to work at the $B_c$ meson rest frame and the
light cone coordinate. The $B_c$ meson momentum $P_1$ and the
charmonium meson  momentum $P_2$ are chosen as \cite{prd67054028}
\begin{eqnarray}
P_1=\frac{M}{\sqrt{2}}(1,1,\textbf{0}_{T}),\quad P_2=\frac{M}{\sqrt{2}}(r\eta^+,r\eta^-,\textbf{0}_{T}),
\end{eqnarray}
with the ratio $r=m/M $ and $m(M)$ is the mass of the charmonium ($B_c$) meson. The factors  $\eta^{\pm}=\eta\pm \sqrt{\eta^2-1}$ come with the definition of the $\eta$  of the form \cite{prd67054028}
\begin{eqnarray}
\eta=\frac{1+r^2}{2r}-\frac{q^2}{2rM^2},
\end{eqnarray}
with the momentum transfer $q=P_1-P_2$. When the final state is a vector meson,  the longitudinal and transverse polarization vector $\epsilon_{L,T}$ can be written as
\begin{eqnarray}
\epsilon_L=\frac{1}{\sqrt{2}}(\eta^+,-\eta^-,\textbf{0}_{T}),\quad
\epsilon_T=(0,0,1).
\end{eqnarray}
The momentum of the valence quarks $k_{1,2}$, whose notation is displayed in Fig. \ref{fig:semi}, is parametrized as
\begin{eqnarray}
k_1=(x_1\frac{M}{\sqrt{2}},x_1\frac{M}{\sqrt{2}},\textbf{k}_{1T}),\quad
k_2=(\frac{M}{\sqrt{2}}x_2r\eta^+,\frac{M}{\sqrt{2}}x_2r\eta^-,\textbf{k}_{2T}),
\end{eqnarray}
the $k_{1T,2T}$,  $x_{1,2}$ represent the transverse
momentum and longitudinal  momentum fraction of the charm quark inside the meson, respectively.
One should note that there is no end-point singularity in the  $B_c$ meson decays and the integral is still convergent without the parton transverse momentum $k_{1T}$ of $B_c$ meson in the collinear factorization. However, we here still keep it to suppress some non-physical contributions near the singularity
(for example the singularity at $x_1=0.1923$ for $B_c\rightarrow J/\psi$ decay).

There are three typical scales of the $B_c$ to
charmonium  decays: $M$, $m$, and  the heavy-meson
and heavy-quark mass difference $\bar{\Lambda}$.
%which possess the hierarchy of $M\gg m \gg \bar{\Lambda}$.
These three scales allow for a consistent power expansion in $m/M$ and in $\bar{\Lambda}/m$ under the hierarchy of $M\gg m \gg \bar{\Lambda}$.
In the heavy-quark and large-recoil limits, based on the $k_T$ factorization theorem,
the corresponding form factors can be expressed as the convolution of the hard amplitude
with $B_c$ and charmonium meson wave functions. The hard amplitude can be treated by perturbative QCD at the leading order in an $\alpha_s$  expansion (single gluon exchange as depicted  in Fig. \ref{fig:semi}).
The higher-order radiative corrections generate the logarithm divergences,
which can be absorbed into the meson wave functions.
One also encounters double logarithm divergences when collinear and soft divergences overlap,
which can be  summed to all orders to give a Sudakov factor.
After absorbing all the soft dynamics, the initial and final state meson wave functions can be treated as  nonperturbative inputs,
which are not calculable but universal.

Similar to the situation of the $B$ meson \cite{npb591313}, under  above  hierarchy, at leading order in $1/M$,
the $B_c$ meson light-cone matrix element can be decomposed as \cite{prd55272}
\begin{eqnarray}
\int d^4 z e^{ik_1 \cdot z}\langle 0|\bar{b}_{\alpha}(0)c(z)_{\beta}|B_c(P_1)\rangle=\frac{i}{\sqrt{2N_c}}
\{(\rlap{/}{P_1}+M)\gamma_5[\Phi_{B_c}(k_1)+\rlap{/}{v}
\bar{\Phi}_{B_c}(k_1)]\}_{\alpha\beta},
\end{eqnarray}
with the unit vectors $v=(0,1,0_T)$ on the light cone.
Here, we only consider the contribution from $\Phi_{B_c}$, while the contribution
of $\bar{\Phi}_{B_c}$ starting from the next-to-leading-power $\bar{\Lambda}/M$ is numerically neglected \cite{epjc28515,prd76074018}.
In coordinate space $\Phi_{B_c}$  can be expressed by
\begin{eqnarray}
\Phi_{B_c}(x)=\frac{i}{\sqrt{2N_c}}[(\rlap{/}{P_1}+M)\gamma_5\phi_{B_c}(x)].
\end{eqnarray}
The distribution amplitude $\phi_{B_c}$ is  adopted in the  form as \cite{prd89114019}
\begin{eqnarray}
\phi_{B_c}(x)=N x(1-x)\exp[-\frac{m_b+m_c}{8m_bm_c \omega}(\frac{m_c^2}{x}+\frac{m_b^2}{1-x})],
\end{eqnarray}
with   shape parameters $\omega=0.5\pm 0.1$ GeV and the normalization conditions
\begin{eqnarray}
\int_0^1\phi_{B_c}(x)d x=1.
\end{eqnarray}
$N$  is  the normalization constant.

For the charmonium meson,
because of its large mass, the higher-twist contributions are important. The light-cone wave functions are obtained in powers of
$m/E$ or $\bar{\Lambda}/E$ where $E(\approx M)$ is the energy of the charmonium meson.
In terms of the notation in
Ref. \cite{prd71114008}, we decompose the nonlocal matrix elements for
the longitudinally and transversely polarized vector mesons ($V=J/\psi, \psi(2S), \psi(3S)$)
and pseudoscalar mesons ($P=\eta_c, \eta_c(2S), \eta_c(3S)$) into
\begin{eqnarray}
\langle V (P_2, \epsilon_L)|\bar{c}(z)_{\alpha}c(0)_{\beta}|0\rangle &=& \frac{1}{\sqrt{2N_c}}\int_0^1 dxe^{ixP_2\cdot z}
[m\rlap{/}{\epsilon_L}_{\alpha\beta}\psi^L(x,b)+(\rlap{/}{\epsilon_L}\rlap{/}{P_2})_{\alpha\beta}\psi^t(x,b)], \nonumber\\
\langle V (P_2, \epsilon_T)|\bar{c}(z)_{\alpha}c(0)_{\beta}|0\rangle &=& \frac{1}{\sqrt{2N_c}}\int_0^1 dxe^{ixP_2\cdot z}
[m\rlap{/}{\epsilon_T}_{\alpha\beta}\psi^V(x,b)+(\rlap{/}{\epsilon_T}\rlap{/}{P_2})_{\alpha\beta}\psi^T(x,b)],\nonumber\\
\langle P (P_2)|\bar{c}(z)_{\alpha}c(0)_{\beta}|0\rangle &=& -\frac{i}{\sqrt{2N_c}}\int_0^1 dxe^{ixP_2\cdot z}
[(\gamma_5\rlap{/}{P_2})_{\alpha\beta}\psi^v(x,b)+m(\gamma_5)_{\alpha\beta}\psi^s(x,b)],\nonumber\\
\end{eqnarray}
respectively.  For the distribution amplitudes  of the 1$S$ and 2$S$ states,
the same form and parameters are adopted as in \cite{prd90114030,epjc75293}. The distribution amplitudes  of the 3$S$ states will be derived in the appendix. %\ref{wavefunction}.
%where were extracted from the correspond Schr$\ddot{o}$dinger states for the harmonic oscillator potential.

\section{ Form factors and semileptonic differential decay rates}\label{sec:cul}
The two factorizable emission Feynman diagrams for the semileptonic $B_c$ decays are given in Fig. \ref{fig:semi}.
The  transition form factors, $F_+(q^2)$, $F_0(q^2)$, $V(q^2)$, and $A_{0,1,2}(q^2)$ are defined via the matrix element \cite{zpc29637},
\begin{eqnarray}\label{eq:form1}
\langle P(P_2)|\bar{c}\gamma^{\mu}b|B_c(P_1)\rangle=[(P_1+P_2)^{\mu}-\frac{M^2-m^2}{q^2}q^{\mu}]F_+(q^2)
+\frac{M^2-m^2}{q^2}q^{\mu}F_0(q^2),
\end{eqnarray}
\begin{eqnarray}\label{eq:form2}
\langle V(P_2)|\bar{c}\gamma^{\mu}b|B_c(P_1)\rangle=\frac{2iV(q^2)}{M+m}\epsilon^{\mu\nu\rho\sigma}
\epsilon^*_{\nu}P_{2\rho}P_{1\sigma},
\end{eqnarray}
\begin{eqnarray}\label{eq:form3}
\langle V(P_2)|\bar{c}\gamma^{\mu}\gamma_5b|B_c(P_1)\rangle&=&2mA_0(q^2)\frac{\epsilon^*\cdot q}{q^2}q^{\mu}+
(M+m)A_1(q^2)[\epsilon^{*\mu}-\frac{\epsilon^*\cdot q}{q^2}q^{\mu}]\nonumber\\&&
-A_2(q^2)\frac{\epsilon^*\cdot q}{M+m}[(P_1+P_2)^{\mu}-\frac{M^2-m^2}{q^2}q^{\mu}],
\end{eqnarray}
with $\epsilon^{0123}=+1$.
In the large-recoil limit ($q^2=0$), the following relations  should  hold to cancel the poles:
\begin{eqnarray}
F_0(0)=F_+(0),\quad A_0(0)=\frac{1+r}{2r}A_1(0)-\frac{1-r}{2r}A_2(0)\;.
\end{eqnarray}
In the  pQCD framework, it is convenient to compute the other equivalent auxiliary form factors $f_1(q^2)$ and $f_2(q^2)$, which are related to $F_+(q^2)$ and $F_0(q^2)$ by \cite{cpc37093102}
\begin{eqnarray}
F_+&=&\frac{1}{2}(f_1+f_2),\nonumber\\
F_0&=&\frac{1}{2}f_1[1+\frac{q^2}{M^2-m^2}]+\frac{1}{2}f_2[1-\frac{q^2}{M^2-m^2}].
\end{eqnarray}
Following the derivation of the factorization formula for the $B\rightarrow P$, $B\rightarrow V$  transitions \cite{prd65014007},
we obtain these form factors as follows:
\begin{eqnarray}\label{eq:f1ex}
f_1(q^2)&=&4 \sqrt{\frac{2}{3}} \pi  M^2 f_B C_f r \int_0^1dx_1dx_2\int_0^{\infty}b_1b_2db_1db_2\phi_{B_c}(x_1)\nonumber\\&&
[\psi ^L(x_2,b_2) r(x_2-1)-\psi ^t(x_2,b_2)(r_b-2)]E_{ab}(t_a)h(\alpha_e,\beta_a,b_1,b_2)S_t(x_2)\nonumber\\&&
-[\psi^L(x_2,b_2)(r-2\eta x_1)+\psi^t(x_2,b_2)2(x_1-r_c)]E_{ab}(t_b)h(\alpha_e,\beta_b,b_2,b_1)S_t(x_1),
\end{eqnarray}
\begin{eqnarray}\label{eq:f2ex}
f_2(q^2)&=&4 \sqrt{\frac{2}{3}} \pi  M^2 f_B C_f  \int_0^1dx_1dx_2\int_0^{\infty}b_1b_2db_1db_2\phi_{B_c}(x_1)\nonumber\\&&
[\psi ^L(x_2,b_2)(2r_b-1-2r\eta (x_2-1))+\psi ^t(x_2,b_2)2r(x_2-1)]E_{ab}(t_a)h(\alpha_e,\beta_a,b_1,b_2)S_t(x_2)\nonumber\\&&
-[\psi^L(x_2,b_2)(r_c+x_1)-\psi^t(x_2,b_2)2r]E_{ab}(t_b)h(\alpha_e,\beta_b,b_2,b_1)S_t(x_1),
\end{eqnarray}
\begin{eqnarray}\label{eq:A0ex}
A_0(q^2)&=&-2 \sqrt{\frac{2}{3}} \pi  M^2 f_B C_f  \int_0^1dx_1dx_2\int_0^{\infty}b_1b_2db_1db_2\phi_{B_c}(x_1)\nonumber\\&&
[\psi ^L(x_2,b_2) \left(1-2 r_b-r(x_2-1)(r-2\eta)\right)-\psi ^t(x_2,b_2)r \left(2 x_2-r_b\right)]E_{ab}(t_a)h(\alpha_e,\beta_a,b_1,b_2)S_t(x_2)\nonumber\\&&
-\psi ^L(x_2,b_2)[r_c+r^2+x_1(1-2r\eta )]E_{ab}(t_b)h(\alpha_e,\beta_b,b_2,b_1)S_t(x_1),
\end{eqnarray}
\begin{eqnarray}\label{eq:A1ex}
A_1(q^2)&=&4 \sqrt{\frac{2}{3}}\frac{r}{1+r} \pi  M^2 f_B C_f\int_0^1dx_1dx_2\int_0^{\infty}b_1b_2db_1db_2\phi_{B_c}(x_1)\nonumber\\&&
[\psi ^V(x_2,b_2) \left(-2 r_b+\eta  r (x_2-1)+1\right)+\psi ^T(x_2,b_2)[\eta  r_b-2 (\eta +r (x_2-1))]]\nonumber\\&&E_{ab}(t_a)h(\alpha_e,\beta_a,b_1,b_2)S_t(x_2)\nonumber\\&&
-\psi ^V(x_2,b_2)[r_c-x_1+\eta r]E_{ab}(t_b)h(\alpha_e,\beta_b,b_2,b_1)S_t(x_1),
\end{eqnarray}
\begin{eqnarray}\label{eq:A2ex}
A_2(q^2)&=&-A_1\frac{(1+r)^2(r-\eta)}{2r(\eta^2-1)}-2 \pi  M^2 f_B C_f\sqrt{\frac{2}{3}}\frac{1+r}{\eta^2-1}
\int_0^1dx_1dx_2\int_0^{\infty}b_1b_2db_1db_2\phi_{B_c}(x_1)\nonumber\\&&
[\psi ^t(x_2,b_2)(r_b (1-\eta  r)+2 r^2 (x_2-1)-2 \eta  r (x_2-2)-2)\nonumber\\&&
-\psi ^L(x_2,b_2)(2 r_b (\eta -r)-\eta +r (\eta  r (x_2-1)-2 \eta ^2 (x_2-1)+x_2))]
\nonumber\\&&E_{ab}(t_a)h(\alpha_e,\beta_a,b_1,b_2)S_t(x_2)\nonumber\\&&
+\psi ^L(x_2,b_2)[r_c (r-\eta )+\eta  r^2+r \left(-2 \eta ^2 x_1+x_1-1\right)+\eta  x_1]
\nonumber\\&&E_{ab}(t_b)h(\alpha_e,\beta_b,b_2,b_1)S_t(x_1),
\end{eqnarray}
\begin{eqnarray}\label{eq:Vex}
V(q^2)&=&2\sqrt{\frac{2}{3}}\pi  M^2 f_B C_f (1+r)\int_0^1dx_1dx_2\int_0^{\infty}b_1b_2db_1db_2\phi_{B_c}(x_1)\nonumber\\&&
[\psi ^V(x_2,b_2) r\left(1-x_2\right)+\psi ^T(x_2,b_2)(r_b-2)]E_{ab}(t_a)h(\alpha_e,\beta_a,b_1,b_2)S_t(x_2)\nonumber\\&&
-\psi ^V(x_2,b_2)rE_{ab}(t_b)h(\alpha_e,\beta_b,b_2,b_1)S_t(x_1),
\end{eqnarray}
with  $r_{b,c}=\frac{m_{b,c}}{M}$.
  $\alpha_e$ and $\beta_{a,b}$ are the virtuality of the internal gluon and quark, respectively. Their expressions are
 \begin{eqnarray}
\alpha_e&=&-M^2[x_1+\eta^+r(x_2-1)][x_1+\eta^-r(x_2-1)],\nonumber\\
\beta_{a}&=&m_b^2-M^2[1+\eta^+r(x_2-1)][1+\eta^-r(x_2-1)],\nonumber\\
\beta_{b}&=&m_c^2-M^2[\eta^+r-x_1][\eta^-r-x_1].
\end{eqnarray}
The explicit
expressions of the  functions $E_{ab}$, the scales $t_{a,b}$,
and the hard functions $h$ are referred
to  \cite{prd90114030}.
In fact, if  we take $q^2\rightarrow 0$,
these expressions are agree with the results in Ref. \cite{epjc75293}.
At the quark level, the charged current  $B_c\rightarrow P(V)l \nu$ decays occur via the $b\rightarrow c l \nu_l$ transition.
The effective Hamiltonian for the $b\rightarrow c l \nu_l$ transition is written as \cite{rmp1125}
\begin{eqnarray}
\mathcal{H}_{eff}=\frac{G_F}{\sqrt{2}}V^*_{cb}\bar{b}\gamma_{\mu}
(1-\gamma_5)c\otimes \bar{\nu_l}\gamma^{\mu}(1-\gamma_5)l,
\end{eqnarray}
where $G_F=1.16637 \times 10^{-5} \mathrm{GeV}^{-2} $ is the Fermi coupling
constant and $V_{cb}$ is one of the CKM matrix elements.

The differential decay rate of $B_c\rightarrow P l \nu$ reads \cite{prd79054012}
\begin{eqnarray}\label{eq:d1}
\frac{d\Gamma}{dq^2}(B_c\rightarrow P l \nu)=\frac{G_F^2|V_{cb}|^2}{384\pi^3M^3q^2}\sqrt{\lambda(q^2)}
(1-\frac{m_l^2}{q^2})^2[3m_l^2(M^2-m^2)^2|F_0(q^2)|^2
+(m_l^2+2q^2)\lambda(q^2)|F_+(q^2)|^2],
\end{eqnarray}
where $m_l$ is the mass of the leptons and $\lambda(q^2)=(M^2+m^2-q^2)^2-4M^2m^2$.
Since electron and muon are very light
compared with the charm quark, we can safely neglect the masses of these two kinds of leptons in the analysis.
%The mass of the light leptons $e$ and $\mu$ can be ignored.
For the channel of $B_c\rightarrow V l \nu$, the decay
rates in the transverse and longitudinal polarization of the
vector charmonium can be formulated as \cite{prd79054012}
\begin{eqnarray}\label{eq:d2}
\frac{d\Gamma_{\pm}}{dq^2}(B_c\rightarrow V l \nu)&=&\frac{G_F^2|V_{cb}|^2}{384\pi^3M^3}\lambda^{3/2}(q^2)
(1-\frac{m_l^2}{q^2})^2(m_l^2+2q^2)[\frac{V(q^2)}{M+m}
\mp\frac{(M+m)A_1(q^2)}{\sqrt{\lambda(q^2)}}]^2,\nonumber\\
\frac{d\Gamma_{L}}{dq^2}(B_c\rightarrow V l \nu)&=&\frac{G_F^2|V_{cb}|^2}{384\pi^3M^3q^2}\sqrt{\lambda(q^2)}
(1-\frac{m_l^2}{q^2})^2\nonumber\\&&
\{3m_l^2\lambda(q^2)A_0^2(q^2)
+\frac{m_l^2+2q^2}{4m^2}[(M^2-m^2-q^2)(M+m)A_1(q^2)-
\frac{\lambda(q^2)}{M+m}A_2(q^2)]^2\}.
\end{eqnarray}
The combined transverse and total
differential decay widths are defined as
\begin{eqnarray}\label{eq:d3}
\frac{d\Gamma_{T}}{dq^2}=\frac{d\Gamma_{+}}{dq^2}+
\frac{d\Gamma_{-}}{dq^2},\quad \frac{d\Gamma}{dq^2}=\frac{d\Gamma_{T}}{dq^2}+
\frac{d\Gamma_{L}}{dq^2}.
\end{eqnarray}

\section{ Numerical results and discussions}\label{sec:res}
In our calculations, some parameters  are used as inputs, which are listed in Table \ref{tab:constant}.
\begin{table}
\caption{The values of the input parameters for numerical analysis. The tensor decay constant $f^{T}_V$ are determined through the assumption $f^T_V m_V=2 f_V m_c$,
which has been used in \cite{prd63074011}. }
\label{tab:constant}
\begin{tabular*}{16.5cm}{@{\extracolsep{\fill}}l|ccccc}
  \hline\hline
\textbf{Mass(\text{GeV})}    & $M_{B_c}=6.277$ \cite{epjc75293}& $m_{b}=4.18$ \cite{epjc75293}
& $m_{c}=1.275$ \cite{epjc75293}& $ m_{\tau}=1.777$ \cite{prd77054003} & $ m_{J/\psi}=3.097$ \cite{prd90114030} \\[1ex]
&$ m_{\eta_c}=2.981$ \cite{prd90114030}& $m_{\psi(2S)}=3.686$ \cite{epjc75293}
& $m_{\eta_c(2S)}=3.639$ \cite{epjc75293}&$ m_{\eta_c(3S)}=3.940$ \cite{prd77054003}& $m_{\psi(3S)}=4.040$ \cite{prd77054003}\\[1ex]
\hline
\end{tabular*}
\begin{tabular*}{16.5cm}{@{\extracolsep{\fill}}l|cc}
  \hline
\multirow{2}{*}{{\textbf{CKM}}} &$V_{cb}=40.9 \times 10^{-3}$ \cite{epjc75293}
& $V_{ud}=0.97425$\cite{epjc75293} \\[1ex]
\hline
\end{tabular*}
\begin{tabular*}{16.5cm}{@{\extracolsep{\fill}}l|cccc}
\hline
\textbf{Decay constants(MeV)} & $f_{B_c}=489$ \cite{prd90114030}& $f_{\pi}=131$ \cite{prd90114030}& $f_{J/\psi}=405\pm 14 $ \cite{prd90114030}& $f_{\eta_c}=420\pm 50$\cite{prd90114030}\\[1ex]
 & $f_{\psi(2S)}=296^{+3}_{-2}$ \cite{epjc75293}& $f_{\eta_c(2S)}=243^{+79}_{-111} $ \cite{epjc75293}& $f_{\psi(3S)}=187\pm 8$ \cite{prd77054003}& $f_{\eta_c(3S)}=180^{+27}_{-32} $\cite{prd77054003}\\[1ex]
  & $f^T_{J/\psi}=333\pm 12 $ & $f^T_{\psi(2S)}=205^{+2}_{-1}$  & $f^T_{\psi(3S)}=118 \pm 5$ \\[1ex]
\hline
\end{tabular*}
\begin{tabular*}{16.5cm}{@{\extracolsep{\fill}}l|l}
\hline
\textbf{Lifetime} & $\tau_{B_c}=0.453\times 10^{-12}\text{s}$\cite{prd90114030} \\[1ex]
\hline\hline
\end{tabular*}
\end{table}

As  known, the pQCD results of these form factors are reliable only in the small $q^2$ region.
For the form factors in the large $q^2$ region, the fast rise of the pQCD results
indicates that the perturbative calculation gradually becomes unreliable. In order to extend our results to the whole
physical region,  we first
perform the pQCD calculations to these form factors in the lower $q^2$ region ($q^2\in (0,\xi(M-m)^2)$ with $\xi=0.2(0.5)$
 for the $B_c\rightarrow 1S(2S/3S)$ transition), and then we
make an extrapolation  for them to the larger $q^2$ region ($q^2\in (\xi(M-m)^2,(M-m)^2)$).
There exist in the literature several different approaches
for extrapolating the form factors from the small $q^2$ region to the large $q^2$ region.
The three-parameter form is one of the pervasive  models, where the fit function is chosen as
 \begin{eqnarray}\label{eq:extr}
\mathcal {F}_i(q^2)=\mathcal {F}_i(0)\exp[a\frac{q^2}{M^2}+b(\frac{q^2}{M^2})^2],
\end{eqnarray}
where $\mathcal {F}_i$ denotes any   of the form factors, and $a$, $b$ are the fitted  parameters.

%With the above input parameters and the relevant formula,
Our results of the transition form factors at the scale  $q^2=0$ together with the
fitted  parameters $a$, $b$ are collected in Table \ref{tab:formfact}, where the theoretical uncertainties are estimated including three aspects. 

The first kind of uncertainties is from  the shape parameters $\omega$ in the initial and final states and the charm-quark mass $m_c$. In the evaluation, we vary the values of $\omega$ within a $20\%$ range and $m_c=1.275$ GeV by $\pm 0.025$ GeV. We find that, in this work, the form factors are less sensitive to
these  hadronic parameters than our previous studies \cite{prd90114030,epjc75293}.
For example, the error induced by $m_c$ is just a few percent here, while in Ref. \cite{epjc75293} this can reach $10-20\%$.
This can be understood from the $B_c$ meson wave function.  In Ref. \cite{epjc75293}, the $\delta$ function depend strongly on the
mass of charm quark which  results in a relative large uncertainty.
The second error comes from the decay constants of the final  charmonium  meson, which are shown  in Table \ref{tab:constant}.
Due to the low accuracy measurement of the decay width of the double photons decay of the pseudoscalar charmonia, the relevant uncertainty
of  $F_{0.+}$ is large. The last one is caused by the variation of the hard scale from 0.75 to 1.25t.  Most of this uncertainty is less than $10\%$, which means the next-to-leading-order contributions can be safely neglected. The errors from the uncertainty of the CKM matrix elements are very small, and they have been neglected.

\begin{figure}[!htbh]
\begin{center}
\vspace{0cm} \centerline{\epsfxsize=12 cm \epsffile{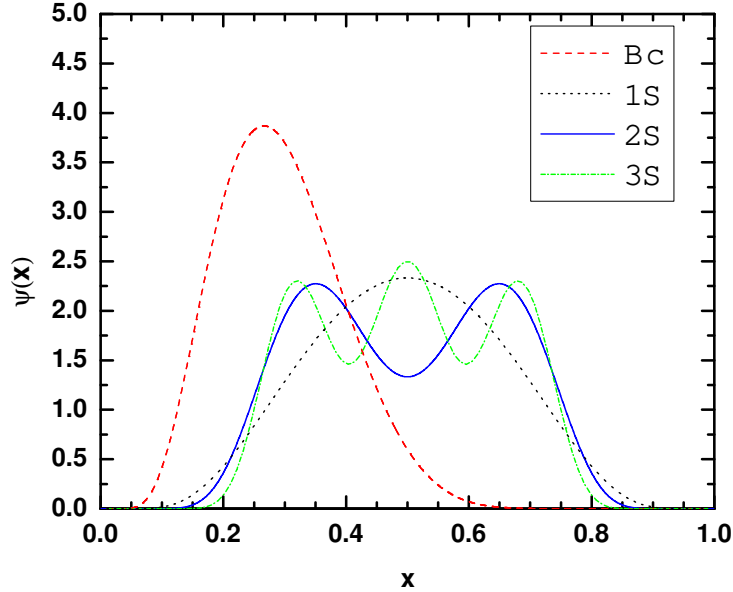}}
\vspace{1.6cm} \caption{The overlap of the leading twist distribution amplitudes of the initial and final state at $b=0$. Dashed, dotted, solid and short dash dotted lines  correspond to $B_c$, 1$S$, 2$S$ and 3$S$ states, respectively.}
 \label{fig:wave}
 \end{center}
\end{figure}

It shows  that the $B_c\rightarrow P/V(1S, 2S)$
transition form factors are a bit larger than our previous calculations \cite{prd90114030,epjc75293}. It is because, here, instead of  the traditional zero-point wave function, we have used the light-cone wave
function for the $B_c$ meson \cite{prd89114019}.  The shape of the leading twist distribution amplitude
of the $B_c$ meson together with the final S-wave  charmonium states are displayed in Fig. \ref{fig:wave}. It is easy to see that the dashed line is broader in shape than that of the zero-point wave function ($\phi_{B_c}(x)\propto \delta(x-r_c)$). The overlap
between the initial and final state wave functions becomes
larger in this work, which certainly induces larger form factors. We also can see that the form factors of the $B_c$ weak transitions
to the 2$S$ charmonium states at zero momentum transfer are comparable with the corresponding values of $B_c$ decays to the 1$S$ charmonium states in  Table \ref{tab:formfact}.
Since one of the peaks of the 2$S$ charmonium states wave function is so close to  the peak of the $B_c$ meson wave function, the overlaps between them are large, which enhances the values of the $B_c\rightarrow  P/V(2S)$ form factors. However, due to the presence of the nodes in the 3$S$ states wave function and the smaller decay constants,
the corresponding form factors of the $B_c$ decays to the 3$S$  states are slightly suppressed.

\begin{table}
\caption{The fit parameters $a$, $b$, and the pQCD predictions of $F_{0,+}(0)$, $A_{0,1,2}(0)$, and $V(0)$ for $B_c\rightarrow nS(n=1,2,3)$ decays,
where the uncertainties come from the
hadronic parameters including shape parameters $\omega$ in  the initial and final state wave functions and charm-quark mass $m_c$,  decay constants, and the hard scale $t$, respectively.}
\label{tab:formfact}
\begin{tabular*}{17.5cm}{l c c c|lccc|lccc}
\hline\hline
$\mathcal {F}_i$&$\mathcal {F}^{B_c\rightarrow 1S}_i(0)$  & a & b &$\mathcal {F}_i$&$\mathcal {F}^{B_c\rightarrow 2S}_i(0)$ & a & b
&$\mathcal {F}_i$&$\mathcal {F}^{B_c\rightarrow 3S}_i(0)$ & a & b \\ \hline
$F_{0}$\;\; & $1.06^{+0.09+0.13+0.10}_{-0.08-0.13-0.02}$ \;\;&3.36 &10.21 &$F_{0}$\;\;& $1.04^{+0.13+0.34+0.13}_{-0.10-0.48-0.03}$ \;\;&4.12 &-30.33&$F_{0}$\;\;& $0.78^{+0.14+0.12+0.08}_{-0.13-0.14-0.02}$ \;\;&1.81 &-167.96 \\
$F_{+}$\;\; &$1.06^{+0.09+0.13+0.10}_{-0.08-0.13-0.02}$  \;\;&4.18 &10.46 &$F_{+}$\;\; & $1.04^{+0.13+0.34+0.13}_{-0.10-0.48-0.03}$ \;\;&5.28 &-26.73&$F_{+}$\;\; & $0.78^{+0.14+0.12+0.08}_{-0.13-0.14-0.02}$  \;\;&3.25&-155.92\\
$A_{0}$\;\; &$0.78^{+0.10+0.03+0.08}_{-0.06-0.02-0.00}$  \;\;&5.41 &10.86  &$A_{0}$\;\; &$0.80^{+0.13+0.01+0.07}_{-0.11-0.01-0.01}$ \;\;&5.16 &-21.08&$A_{0}$\;\; &$0.41^{+0.10+0.02+0.04}_{-0.09-0.02-0.01}$  \;\;&-3.01 &-98.48\\
$A_{1}$\;\; &$0.96^{+0.11+0.04+0.10}_{-0.07-0.03-0.01}$  \;\;&5.24 &-15.18 &$A_{1}$\;\; &$0.87^{+0.17+0.01+0.10}_{-0.11-0.01-0.00}$ \;\;&3.23 &-25.03&$A_{1}$\;\; &$0.41^{+0.08+0.02+0.03}_{-0.08-0.02-0.00}$  \;\;&-3.07 &-162.03\\
$A_{2}$\;\; &$1.36^{+0.16+0.04+0.17}_{-0.12-0.06-0.00}$  \;\;&7.60 &-5.94  &$A_{2}$\;\; &$1.22^{+0.28+0.02+0.22}_{-0.10-0.02-0.00}$ \;\;&8.51 &-63.77& $A_{2}$\;\; &$0.66^{+0.05+0.03+0.01}_{-0.11-0.03-0.01}$  \;\;&0.10&-19.12\\
$V$\;\; &$1.59^{+0.11+0.06+0.14}_{-0.16-0.05-0.02}$  \;\;&5.04 &5.88 &$V$\;\; &$1.71^{+0.47+0.02+0.13}_{-0.23-0.02-0.05}$ \;\;&3.77 &-3.78   &$V$\;\; &$1.07^{+0.20+0.05+0.09}_{-0.18-0.05-0.02}$  \;\;&0.69 &-116.48\\
\hline\hline
\end{tabular*}
\end{table}

\begin{figure}
\centering
 \centerline{\epsfxsize=9 cm
 \subfigure{\epsffile{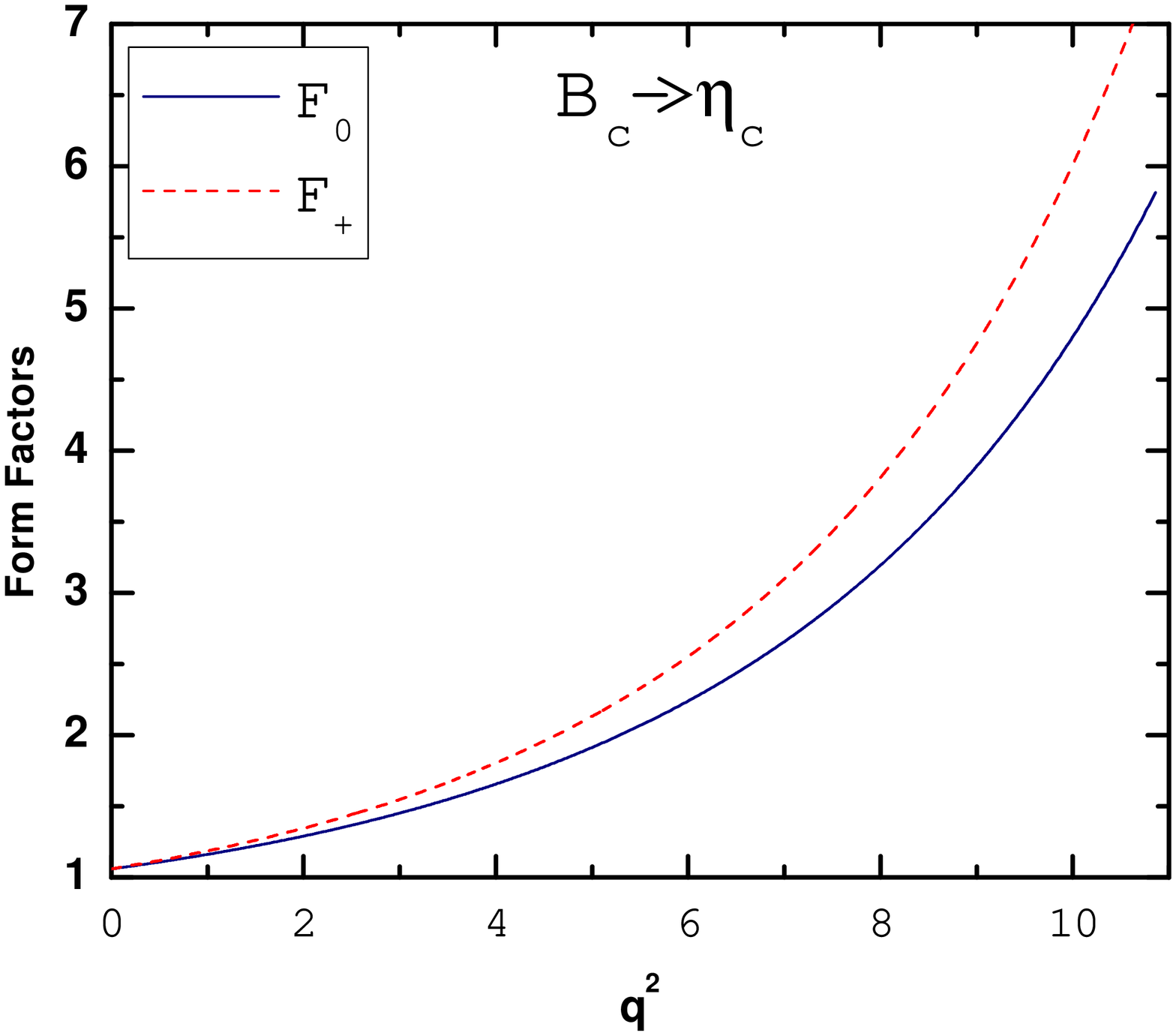}}
\subfigure{\epsffile{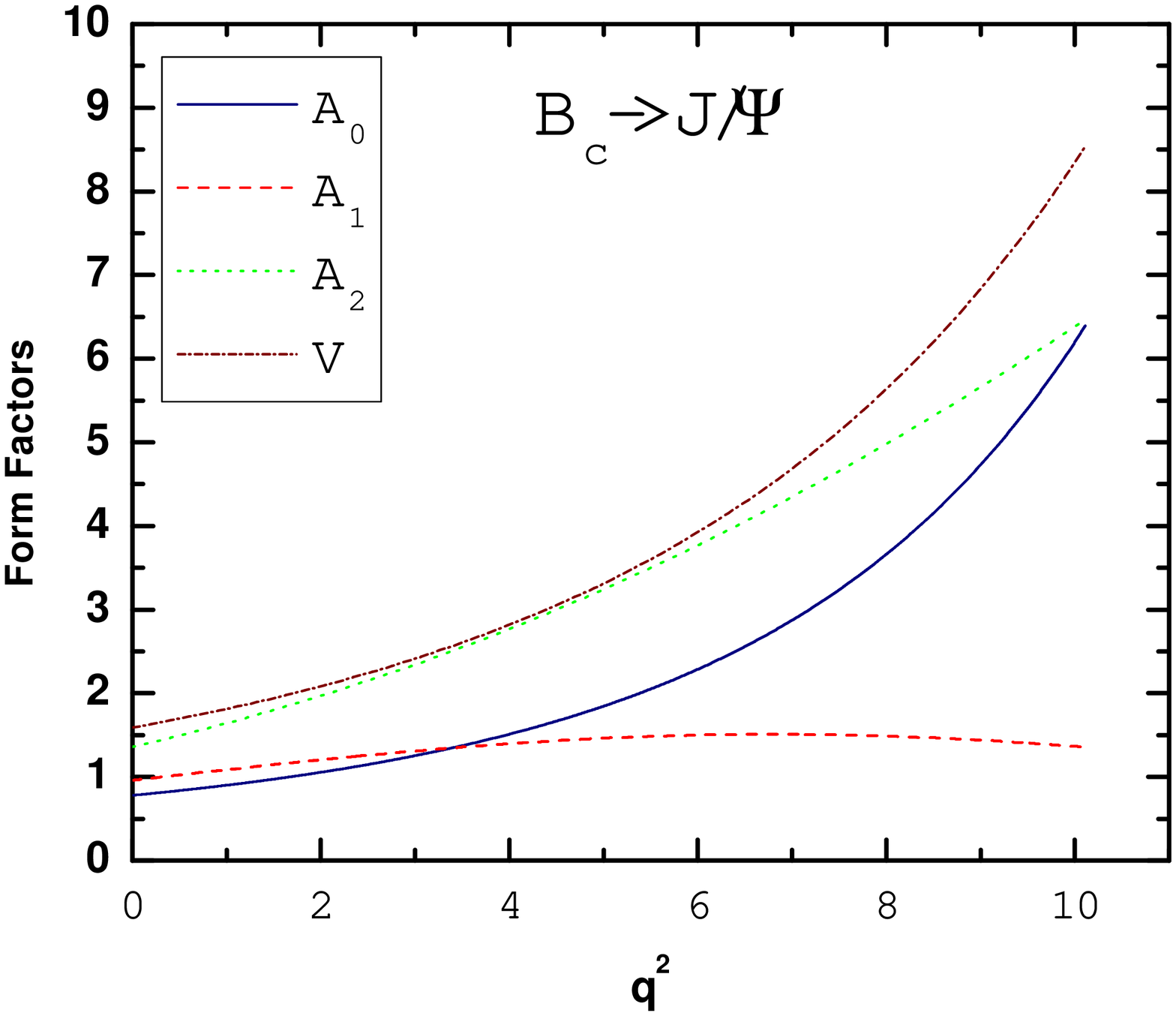}}}
 \centerline{\epsfxsize=9 cm
 \subfigure{\epsffile{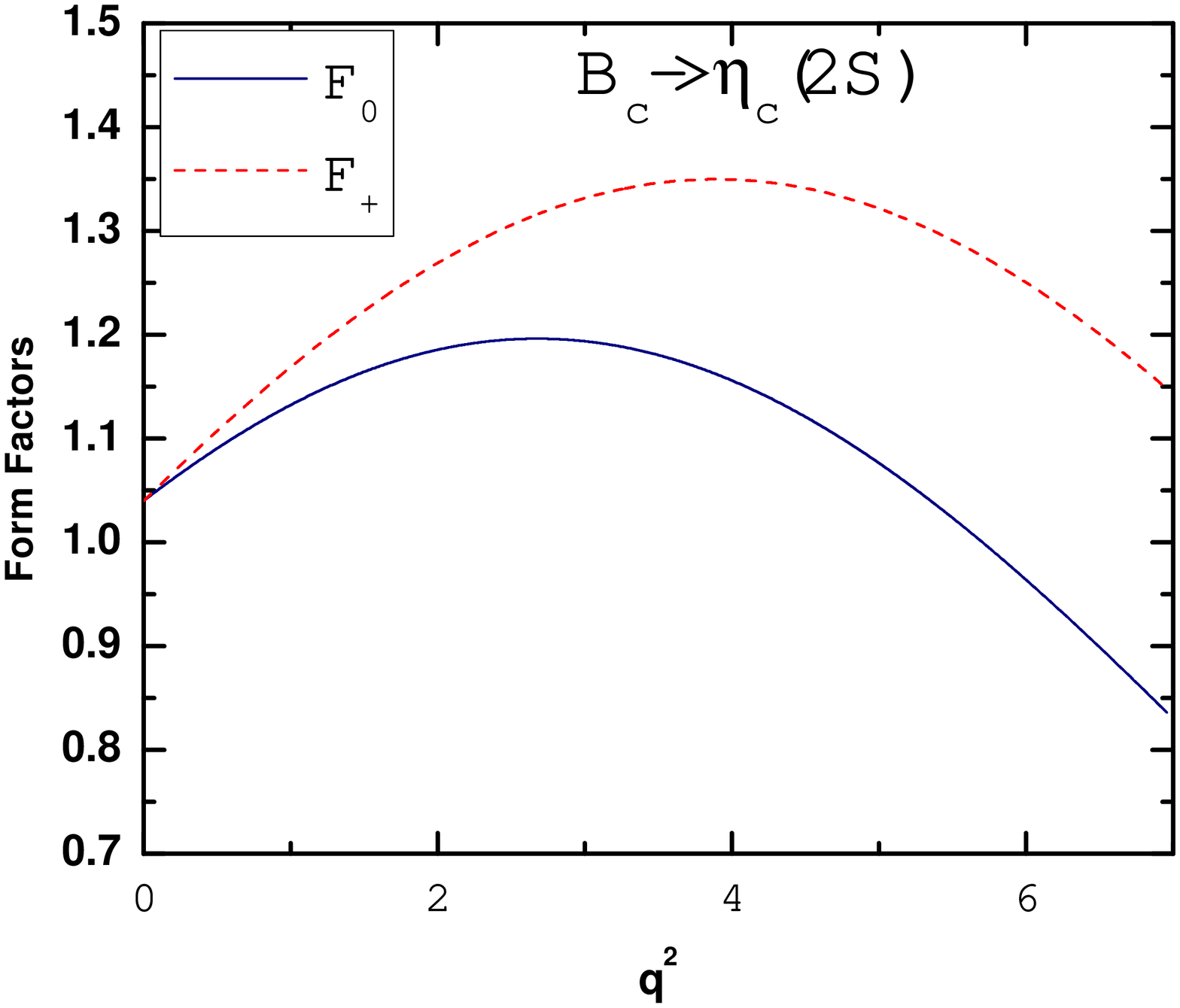}}
\subfigure{\epsffile{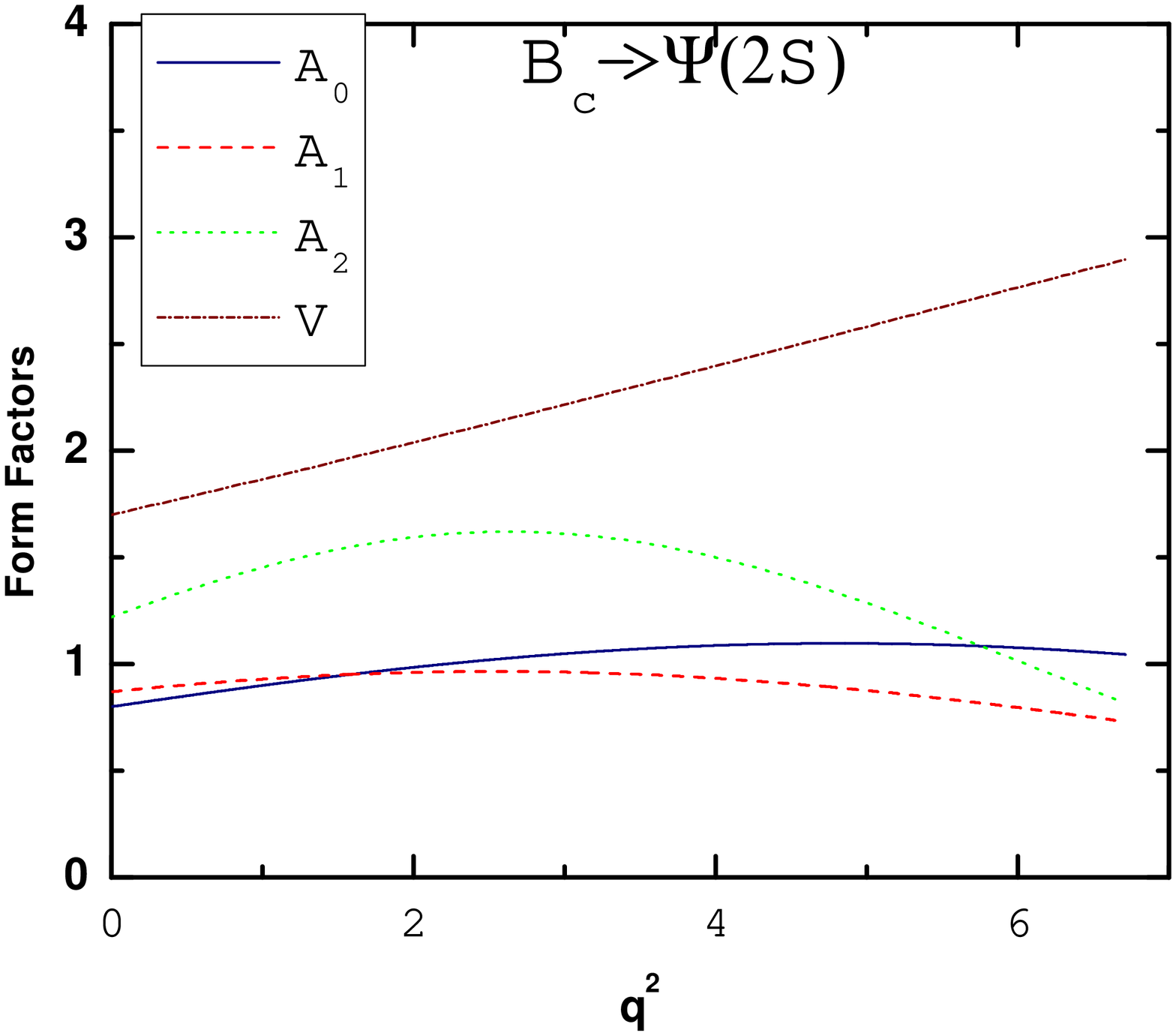}}}
 \centerline{\epsfxsize=9 cm
 \subfigure{\epsffile{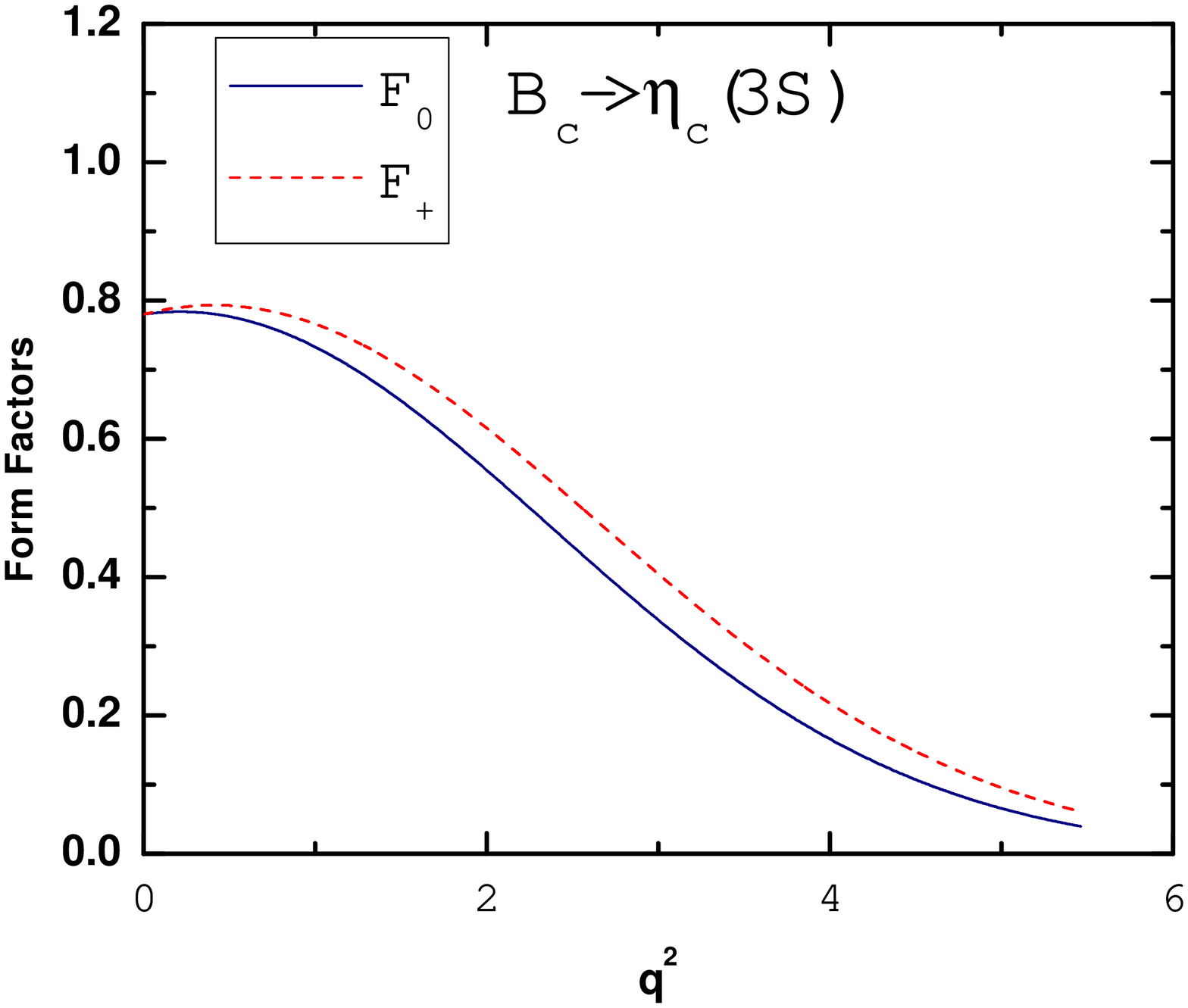}}
\subfigure{\epsffile{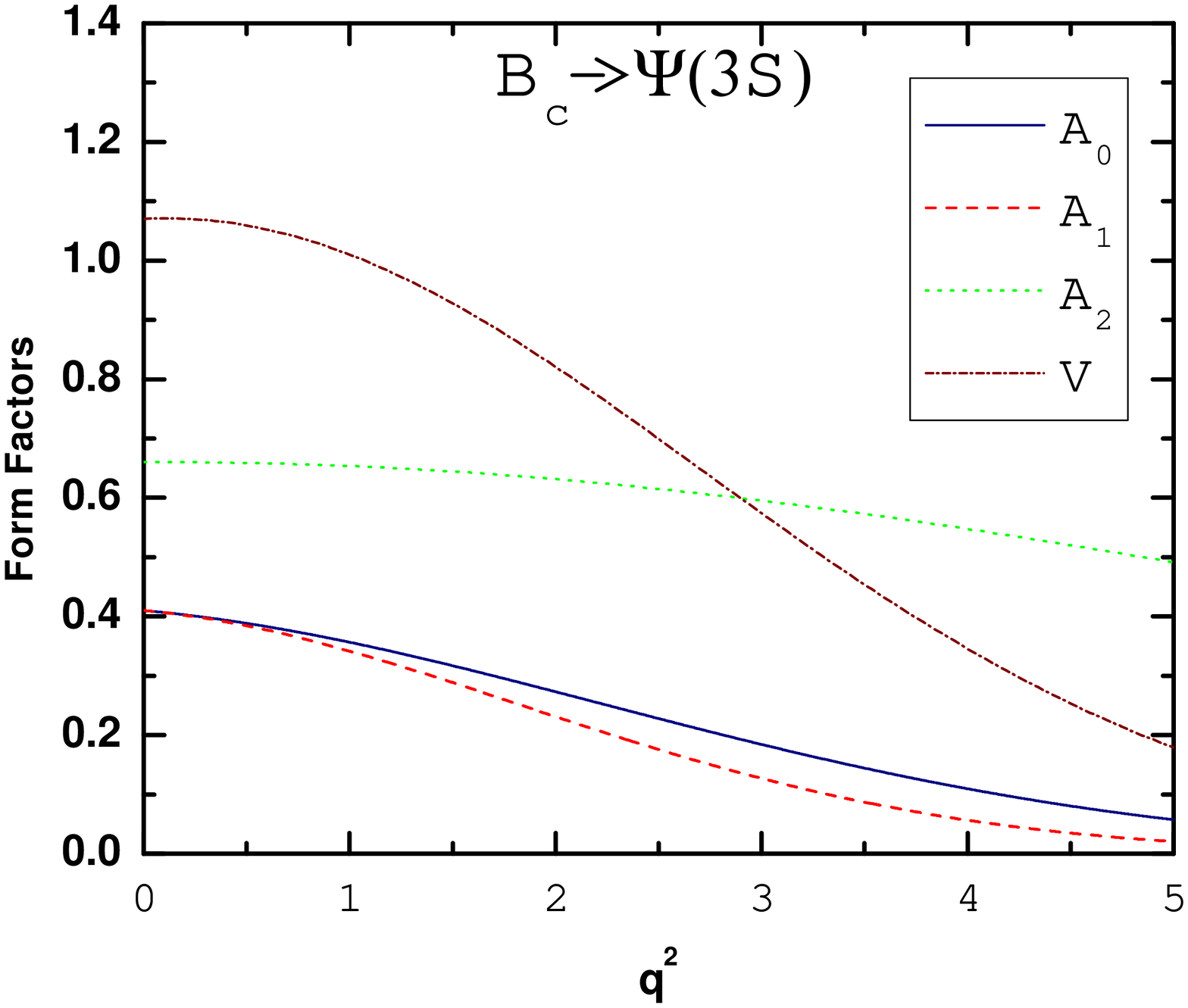}}}
%\subfigure{\includegraphics{fig4c.eps}}
%\subfigure{\includegraphics{fig4d.eps}}
\caption{ Form factors of the $B_c$ decays to the S-wave charmonium states  defined as in Eq.(\ref{eq:extr}).
The left panel is for the $B_c\rightarrow P$ processes, while the right panel for $B_c\rightarrow V$ processes.} \label{fig:form}
\end{figure}
We plot the $q^2$ dependence of the
weak form factors with
center values without theoretical uncertainties in Fig. \ref{fig:form} for
the six decay processes
 in their physical kinematic range.
 We can see the different $q^2$ dependence of the form factors among the $B_c$ decays to different S-wave charmonia clearly. For example,  the form factors
 for the $B_c\rightarrow P/V(1S)$ transition have a relatively strong  $q^2$ dependence, but those of the $B_c\rightarrow P/V(2S/3S)$ transition show a little weaker $q^2$ dependence. In addition, most of these form factors become larger with increasing $q^2$. However,
 this behavior is not universal. For instance, from Fig. \ref{fig:form} some of the form factors for $B_c\rightarrow P/V(2S/3S)$ decays decreases with the increasing $q^2$ in the large region. A similar situation also exists in the light-front quark model \cite{prd89017501} and in the ISGW2 quark model \cite{11022190}.
This behavior of the difference for the corresponding final states is the consequence of their different nodal structure in the wave functions.

\begin{table}
\caption{Branching ratios (in units of \%)  of $B_c\rightarrow P/Vl\nu_l$ decays evaluated by pQCD and by other methods in the literature.
The errors induced by the same sources as  in Table \ref{tab:formfact}.}
\label{tab:br}
\begin{tabular}[t]{l c c c c c cccccc}
\hline\hline
%\toprule[2pt]
Modes     & This work & \cite{prd87014009}&\cite{14113428}&\cite{prd68094020} &\cite{prd63074010}&\cite{prd77054003}&\cite{epjc51833}&\cite{prd79054012}&\cite{prd74074008}&\cite{prd61034012}&\cite{11022190}
 \\ \hline
$B^+_c\rightarrow \eta_c e^+\nu_e$  & $4.5^{+0.7+1.2+0.9}_{-0.7-1.0-0.2}$    &  2.1 &0.55&0.42 &0.81&  &1.64 &0.67 &0.48&0.5&\\
$B^+_c\rightarrow J/\psi e^+\nu_e$  & $5.7^{+1.2+0.5+1.1}_{-0.8-0.4-0.2}$    &   6.7 &1.73  &1.23&2.07& &2.37 &1.49&1.54&3.3&\\
$B^+_c\rightarrow \eta_c \tau^+\nu_{\tau}$  & $2.8^{+0.4+0.7+0.6}_{-0.4-0.6-0.1}$& 0.64 &&& 0.22& &0.49 &0.19&0.17&&\\
$B^+_c\rightarrow J/\psi \tau^+\nu_{\tau}$  & $1.7^{+0.4+0.1+0.3}_{-0.3-0.1-0.1}$&  0.52 &&&0.49 & &0.65 &0.37&0.41&&\\ \hline

$B^+_c\rightarrow \eta_c(2S) e^+\nu_e$  & $0.77^{+0.20+0.58+0.20}_{-0.14-0.55-0.05}$  &  &0.07   &0.03   &  &0.11   &   && &0.02   &0.046                  \\
$B^+_c\rightarrow \psi(2S) e^+\nu_e$  & $1.2^{+0.6+0.1+0.3}_{-0.3-0.1-0.1}$     & &0.1    &0.03   &   &   &   &  &&0.12  &  0.21           \\
$B^+_c\rightarrow \eta_c(2S) \tau^+\nu_{\tau}$  & $5.3^{+1.4+4.1+1.4}_{-1.0-3.8-0.3}\times 10^{-2}$  & &  &  &&$0.81\times10^{-2}$  &&  &  &&$1.3\times 10^{-3}$        \\
$B^+_c\rightarrow \psi(2S) \tau^+\nu_{\tau}$  & $8.4^{+3.6+0.4+1.5}_{-1.3-0.4-0.1}\times 10^{-2}$   & & &&  &  &  & &&  &$1.5\times 10^{-2}$         \\ \hline
$B^+_c\rightarrow \eta_c(3S) e^+\nu_e$  & $0.14^{+0.05+0.05+0.03}_{-0.04-0.04-0.01}$   & &  & $5.5\times10^{-4}$ &  &$1.9\times 10^{-2}$ &  & & &                  \\
$B^+_c\rightarrow \psi(3S) e^+\nu_e$  & $3.6^{+1.8+0.3+0.6}_{-1.3-0.4-0.0}\times 10^{-2}$    & &  & $5.7\times10^{-4}$   &  & & & &    &        \\
$B^+_c\rightarrow \eta_c(3S) \tau^+\nu_{\tau}$  & $1.9^{+0.7+0.6+0.4}_{-0.6-0.6-0.1}\times 10^{-4}$    & & &$5.0\times10^{-7}$ & &$5.7\times 10^{-4}$ & &  &&             \\
$B^+_c\rightarrow \psi(3S) \tau^+\nu_{\tau}$  & $3.8^{+1.5+0.2+0.5}_{-1.4-0.4-0.1}\times 10^{-5}$   & & & $3.6\times10^{-6}$  &  &  & &  & &               \\
%\bottomrule[2pt]
\hline\hline
\end{tabular}
%\footnotetext[1]{We quote the result with the modified wave functions  for $\psi(2S)$.}
\end{table}
Integrating the expressions in Eqs. (\ref{eq:d1}) and   (\ref{eq:d2}) over the variable
$q^2$ in the physical kinematical region, one obtains the relevant decay widths. Then it is
straightforward to calculate the branching ratios.
The results of our evaluation of the branching ratios for all
the considered decays  appear in Table \ref{tab:br} in comparison with predictions
of other approaches. For the $B_c\rightarrow P/V(1S)$ decays, our results are comparable
to those of \cite{prd87014009} within the error bars, but larger than the
results from other models due to the values of the weak form factors.

 For the $B_c\rightarrow P/V(2S,3S)$ decays our predictions are generally
close to the light-cone QCD sum rules  results of \cite{prd77054003}.
However, the  relativistic quark model predictions for the  $B_c\rightarrow P/V(3S)$ decays  in Refs. \cite{prd68094020}
are typically smaller, which can be discriminated by the future LHC experiments.

From Table \ref{tab:br}, we can see the
former four processes have a relatively large branching
ratio ($10^{-2}$), while the branching ratios of the last four processes are comparatively
small ($10^{-7}\sim 10^{-3}$).  They have the following hierarchy:
 \begin{eqnarray}
\mathcal {BR}(B_c\rightarrow P/V (3S))<\mathcal {BR}(B_c\rightarrow P/V (2S))<\mathcal {BR}(B_c\rightarrow P/V(1S)).
\end{eqnarray}
This is  due to  the tighter phase space, smaller decay constants, and the less sensitive dependence of the form factors on the momentum
transfer $q^2$ for the higher excited  state, which can be seen in Fig. \ref{fig:form}.
 The combined effect above suppresses the branching ratios of the semileptonic $B_c$ decays to radially excited charmonia.
 For decays to higher charmonium excitations such a suppression should be  more pronounced.
 In order to reduce  the theoretical uncertainties from the  hadronic
 parameters and the decay constants, we defined six ratios between the electron and tau branching ratios, i.e.
  \begin{eqnarray}\label{eq:ratio1}
\mathcal {R}(P/V)=\frac{\mathcal {BR}(B_c^+\rightarrow P/V e^+ \nu_e)}{\mathcal {BR}(B_c^+\rightarrow P/V \tau^+\nu_{\tau})}.
\end{eqnarray}
From our numerical values listed in Table \ref{tab:br}, we  obtain
  \begin{eqnarray}
\mathcal {R}(J/\psi)& =&3.4^{+0.1}_{-0.1},\quad
\mathcal {R}(\psi(2S)) =14.3^{+0.9}_{-1.4},\quad
\mathcal {R}(\psi(3S)) =947.4^{+71.1}_{-0.0},\nonumber\\
\mathcal {R}(\eta_c)& =&1.6^{+0.0}_{-0.0},\quad
\mathcal {R}(\eta_c(2S)) =14.5^{+0.0}_{-0.3},\quad
\mathcal {R}(\eta_c(3S)) =736.8^{+21.8}_{-9.5},
\end{eqnarray}
where the errors correspond to the combined uncertainty in the hadronic parameters, decay constants, and the hard scale.
Since these parameter dependences canceled out in Eq. (\ref{eq:ratio1}),
the total theoretical errors of these ratios are only a few percent, much smaller than those for the branching ratios.
In general, these ratios are of the same order of magnitude in the different approaches except the light-cone QCD sum rules \cite{prd77054003}, where it is obtained the smallest values of $\mathcal {R}(\eta_c(3S))=33.3$.

For a more direct comparison with the available experimental data \cite{prd90032009},
we need to recalculate some of the nonleptonic $B_c$ decays by using the same wave functions and input parameters as this paper,
whose results are
 \begin{eqnarray}
\mathcal {BR}(B^+_c\rightarrow J/\psi \pi^+)&=&2.6^{+0.6+0.2+0.8}_{-0.4-0.2-0.2}\times 10^{-3}, \quad
\mathcal {BR}(B^+_c\rightarrow \eta_c \pi^+)=5.2^{+1.3+1.3+1.8}_{-0.6-1.2-0.2} \times 10^{-3},\nonumber\\
\mathcal {BR}(B^+_c\rightarrow \psi(2S) \pi^+)&=&8.2^{+2.1+0.2+2.7}_{-2.4-0.1-0.7} \times 10^{-4}, \quad
\mathcal {BR}(B^+_c\rightarrow \eta_c(2S) \pi^+)=1.3^{+0.5+0.9+0.7}_{-0.1-0.9-0.0} \times 10^{-3},\nonumber\\
\mathcal {BR}(B^+_c\rightarrow \psi(3S) \pi^+)&=&4.8^{+2.0+0.5+1.5}_{-1.7-0.5-0.3} \times 10^{-4}, \quad
\mathcal {BR}(B^+_c\rightarrow \eta_c(3S) \pi^+)=1.4^{+0.4+0.4+0.4}_{-0.5-0.4-0.1} \times 10^{-3},
\end{eqnarray}
where the errors induced by the same sources as  in Table \ref{tab:formfact}.
\begin{table}
\caption{Some of the ratios among the branching fractions of the $B_c$ decays in comparison with the data and other theoretical estimates, Here $l$ stands for $l=e,\mu$. The errors correspond to the combined uncertainty in the hadronic parameters, decay constants, and the hard scale.}
\label{tab:ratio}
\begin{tabular}[t]{l c c c c ccc}
\hline\hline
Ratios     & This work  & NRQCD \cite{prd87014009} & BSRQM \cite{14113428}
 & RQM \cite{prd68094020}& QCDPM \cite{prd61034012} &LFQM \cite{prd89017501}&Data \cite{prd90032009,prd92072007}\\ \hline
$\frac{\mathcal {BR}(B_c^+\rightarrow J/\psi \pi^+)}{\mathcal {BR}(B_c^+\rightarrow J/\psi l^+\nu_{l})} $ &$0.046^{+0.003}_{-0.002}$ &0.043 &0.064&0.050&0.039&0.058&0.0469\\\hline
$\frac{\mathcal {BR}(B_c^+\rightarrow \psi(2S) \pi^+)}{\mathcal {BR}(B_c^+\rightarrow \psi(2S) l^+\nu_{l})} $ &$0.068^{+0.000}_{-0.007}$ & & 0.258&0.355&0.158&0.148\\\hline
$\frac{\mathcal {BR}(B_c^+\rightarrow \eta_c \pi^+)}{\mathcal {BR}(B_c^+\rightarrow \eta_c l^+\nu_{l})} $ &$0.116^{+0.010}_{-0.001}$ & 0.247 &0.191 &0.202&0.052&\\\hline
$\frac{\mathcal {BR}(B_c^+\rightarrow \eta_c(2S) \pi^+)}{\mathcal {BR}(B_c^+\rightarrow \eta_c(2S)l^+\nu_{l})} $ &$0.169^{+0.031}_{-0.000}$& &0.432 &0.531&0.33&\\\hline
$\frac{\mathcal {BR}(B_c^+\rightarrow \psi(2S) \pi^+)}{\mathcal {BR}(B_c^+\rightarrow J/\psi \pi^+)} $ &$0.32^{+0.01}_{-0.04}$ &0.26 &0.20 &0.18&0.15&0.23&0.268\\\hline
$\frac{\mathcal {BR}(B_c^+\rightarrow \eta_c(2S) \pi^+)}{\mathcal {BR}(B_c^+\rightarrow \eta_c \pi^+)} $ &$0.25^{+0.07}_{-0.14}$ & &0.27& 0.20&0.25&\\
\hline\hline
\end{tabular}
\end{table}
The ratios among the branching fractions are shown
explicitly in Table \ref{tab:ratio}, from which we can see that the ratios
$\frac{\mathcal {BR}(B_c^+\rightarrow J/\psi \pi^+)}{\mathcal {BR}(B_c^+\rightarrow J/\psi l^+\nu_{l})} $
and $\frac{\mathcal {BR}(B_c^+\rightarrow \psi(2S) \pi^+)}{\mathcal {BR}(B_c^+\rightarrow J/\psi \pi^+)} $
are well consistent with the recent data  \cite{prd90032009,prd92072007}, and also comparable with the  prediction of the NRQCD  \cite{prd87014009}.
Furthermore the latter still agree with the  previous pQCD calculations \cite{epjc75293} 0.29,
although both  $\mathcal {BR}(B_c^+\rightarrow \psi(2S) \pi^+)$ and $\mathcal {BR}(B_c^+\rightarrow J/\psi \pi^+) $ are enhanced compared
with the corresponding values of  \cite{prd90114030,epjc75293}.

We now investigate the relative importance of  the   longitudinal ($\Gamma_L$) and transverse ($\Gamma_T$)
polarizations contributions
%in Eq. \ref{eq:d2} and Eq. \ref{eq:d3}
to the branching ratios of $B_c\rightarrow V l\nu_l$ decays within Region (1), Region (2), and the whole physical region, whose results and the ratios $\frac{\Gamma_L}{\Gamma_T}$ are displayed separately  in Table \ref{tab:relative}. For light electron and  muon, the regions  are defined as: Region (1): $0<q^2<(M-m)^2/2$; Region (2): $(M-m)^2/2<q^2<(M-m)^2$. For the heavy lepton $\tau$, The first region is $m^2_{\tau}<q^2<[(M-m)^2+m^2_{\tau}]/2$ while the second region is $[(M-m)^2+m^2_{\tau}]/2<q^2<(M-m)^2$. From  Table \ref{tab:relative}, all of $\frac{\Gamma_L}{\Gamma_T}$ are $<1$ in Region (2),  which means that the transverse  polarization dominates the branching ratios in this region.
It can be understood as follows.   For the $B_c\rightarrow 1S,2S$ decays, the form factor $V$ as shown in Fig. \ref{fig:form} increase as the $q^2$ increase, which enhances the transverse polarization contribution in the large $q^2$ region, while for the $B_c\rightarrow 3S$ decay, although the value of  $V$ decreases gradually with   increasing $q^2$,  the form factor $A_1$, which gives a dominant contribution to $\Gamma_L$, is  significantly suppressed in the large region, and as a  results  the dominant contributions
to the branching ratios of $B_c\rightarrow \psi(2S)$ decays come from Region (1).

 For $B_c\rightarrow \psi(2S,3S)e\nu_e$ decays $\Gamma_L$ is comparable with $\Gamma_T$ in the whole physical region. These results will be tested by LHCb and the forthcoming Super-B experiments.
\begin{table}
\caption{ The  partial branching ratios and polarizations $\frac{\Gamma_L}{\Gamma_T}$ of $B_c\rightarrow V l\nu_l$ decays  in different $q^2$ regions.}
%Polarizations of $\frac{\Gamma_L}{\Gamma_T}$ for $B_c\rightarrow V l\nu_l$, together with results from RCQM.
% The errors correspond to the combined uncertainty in the hadronic parameters, decay constants and  the hard scale. }
\label{tab:relative}
\begin{tabular}[t]{lccc|lccc}
\hline\hline
%Bins of $q^2$ (GeV$^2$)
% & $0<q^2<\frac{(M+m)^2}{2}$ & $\frac{(M+m)^2}{2}<q^2<(M+m)^2$ &The whole $q^2$ region\\
 & Region (1)&Region (2) &Total & & Region (1)&Region (2) &Total  \\
\hline
$\mathcal {BR}(B_c^+\rightarrow J/\psi e^+\nu_{e})$ &$2.3\times 10^{-2}$  &$3.4\times 10^{-2}$ &$5.7\times 10^{-2}$
 &$\mathcal {BR}(B_c^+\rightarrow J/\psi \tau^+\nu_{\tau})$ &$0.6\times 10^{-2}$  &$1.1\times 10^{-2}$ &$1.7\times 10^{-2}$ \\
$\frac{\Gamma_L}{\Gamma_T}$ &0.82&0.33&0.49&$\frac{\Gamma_L}{\Gamma_T}$ &0.76&0.57&0.63\\
$\mathcal {BR}(B_c^+\rightarrow \psi(2S) e^+\nu_{e})$ &$6.1\times 10^{-3}$  &$5.5\times 10^{-3}$ &$11.6\times 10^{-3}$
&$\mathcal {BR}(B_c^+\rightarrow \psi(2S) \tau^+\nu_{\tau})$ &$3.1\times 10^{-4}$  &$5.3\times 10^{-4}$ &$8.4\times 10^{-4}$   \\
$\frac{\Gamma_L}{\Gamma_T}$ &1.22&0.56&0.85&$\frac{\Gamma_L}{\Gamma_T}$ &0.82&0.61&0.69\\
$\mathcal {BR}(B_c^+\rightarrow \psi (3S) e^+\nu_{e})$ &$3.3\times 10^{-4}$  &$0.3\times 10^{-4}$ &$3.6\times 10^{-4}$
&$\mathcal {BR}(B_c^+\rightarrow \psi (3S) \tau^+\nu_{\tau})$ &$2.2\times 10^{-7}$  &$1.6\times 10^{-7}$ &$3.8\times 10^{-7}$   \\
$\frac{\Gamma_L}{\Gamma_T}$ &1.38&0.23&1.17&$\frac{\Gamma_L}{\Gamma_T}$ &0.49&0.39&0.45\\
% & $m_{\tau}^2<q^2<\frac{(M+m)^2+m_{\tau}^2}{2}$ & $\frac{(M+m)^2+m_{\tau}^2}{2}<q^2<(M+m)^2$ &The whole $q^2$ region\\ \hline
%$\mathcal {BR}(B_c^+\rightarrow J/\psi \tau^+\nu_{\tau})$ &$2.3\times 10^{-2}$  &$3.4\times 10^{-2}$ &$5.7\times 10^{-2}$   \\
%$\frac{\Gamma_L}{\Gamma_T}$ &0.82&0.33&0.49\\
%$\mathcal {BR}(B_c^+\rightarrow \psi(2S) \tau^+\nu_{\tau})$ &$5.3\times 10^{-3}$  &$1.9\times 10^{-3}$ &$7.2\times 10^{-3}$   \\
%$\frac{\Gamma_L}{\Gamma_T}$ &1.36&0.45&1.04\\
%$\mathcal {BR}(B_c^+\rightarrow \psi (3S) \tau^+\nu_{\tau})$ &$1.9\times 10^{-4}$  &$2.3\times 10^{-4}$ &$4.2\times 10^{-4}$   \\
%$\frac{\Gamma_L}{\Gamma_T}$ &2.02&0.78&1.18\\
\hline\hline
\end{tabular}
\end{table}

\section{ conclusion}
We calculate the transition form factors and obtain the branching ratios of the
semileptonic  decays of $B_c$ meson to S-wave charmonium states  by employing
the pQCD factorization approach.  By using the light-cone wave function for the  $B_c$ meson,
the theoretical uncertainties  from the nonperturbative
 hadronic parameters are largely reduced.
It is found that the processes  of $B_c$ to the ground state charmonium   have   comparatively large branching ratios ($10^{-2}$), while the branching
ratios of other  processes are relatively small owing to  the  phase space suppression, smaller decay constants, and the weaker  $q^2$ dependence of the  form factors.
The theoretically evaluated ratio $\frac{\mathcal {BR}(B_c^+\rightarrow J/\Psi \pi^+)}{\mathcal {BR}(B_c^+\rightarrow J/\Psi \mu^+\nu_{\mu})}=0.046^{+0.003}_{-0.002}$
is consistent with  the recent data from  LHCb.
In addition, some interesting ratios  among these branching fractions are discussed and compared  with  other studies.
In general, these ratios in the different approaches are of the same order of
magnitude, while there are also large discrepancies  for specific decay modes.
%which could be tested at the ongoing and forthcoming experiments.
The partial branching ratios for transverse and longitudinal polarizations
were investigated separately in $B_c \rightarrow V l\nu_l$ decays. We  found that the  transverse polarization gives a large contribution in the large $q^2$ region.
For the semileptonic $B_c\rightarrow \psi(2S,3S)e\nu_e$ decays  the longitudinal  contribution is comparable with the transverse contribution in the whole physical region.
These theoretical  predictions could be tested at the ongoing and forthcoming experiments.

\begin{acknowledgments}
The authors are grateful to Wen-Fei Wang and  Ying-Ying Fan
for helpful discussions. This work is supported in part by the  National Natural Science Foundation of China under
Grants No. 11547020 and No. 11605060, in part by the Natural Science Foundation of
Hebei Province  under Grant No. A2014209308, in part by the Program for the Top Young Innovative Talents of Higher Learning Institutions of Hebei Educational Committee under Grant No. BJ2016041, and in part by the Training Foundation of  North China University of Science and Technology  under Grant No. GP201520 and No. JP201512.
%Program for the Top Young Innovative Talents of Higher Learning Institutions of Hebei Province
\end{acknowledgments}

\begin{appendix}
\section{Wave functions of the 3S states}\label{wavefunction}
In the quark model, $\eta_c(3S)$
and $\psi(3S)$ are the second excited states of $\eta_c$ and $J/\psi$, respectively.
The 3$S$ means that, for these states, we have the radial quantum
number $n = 3$ and the orbital angular momentum $l=0$.
The radial wave function of the corresponding Schr$\ddot{o}$dinger state for the
harmonic-oscillator potential is given by
\begin{eqnarray}
\Psi_{(3S)}(\textbf{r})\propto(\frac{15}{4}-5\alpha^2r^2+\alpha^4 r^4)e^{-\frac{\alpha^2r^2}{2}},
\end{eqnarray}
where $\alpha^2=\frac{m\omega}{2}$ and $\omega$ is the frequency of oscillations or the quantum of
energy.  We perform the Fourier transformation  to the momentum space to get $\Psi_{3S}({\bf k})$ as
\begin{eqnarray}
\Psi_{(3S)}(\textbf{k})\propto(15\alpha^4-20 \alpha^2 k^2+4 k^4)e^{-\frac{k^2}{2\alpha^2}},
\end{eqnarray}
with $k^2$ being the square of the three momentum. In terms of the substitution assumption,
\begin{eqnarray}
\textbf{k}_{\perp}\rightarrow \textbf{k}_{\perp},\quad k_z\rightarrow \frac{m_0}{2}(x-\bar{x}),
\quad m_0^2=\frac{m_c^2+\textbf{k}^2_{\perp}}{x\bar{x}},
\end{eqnarray}
with $m_c$ the $c$-quark mass and $\bar{x}=1-x$. We should make the following replacement as regards the
variable $k^2$
\begin{eqnarray}
k^2\rightarrow \frac{\textbf{k}^2_{\perp}+(x-\bar{x})^2m_c^2}{4x\bar{x}}.
\end{eqnarray}
Then the wave function
can be taken as
\begin{eqnarray}
\Psi_{(3S)}(\textbf{k})\rightarrow \Psi_{(3S)}(x,\textbf{k}_{\perp})
\propto[15\alpha^4-\frac{5\alpha^2(\textbf{k}^2_{\perp}+m_c^2(x-\bar{x})^2)}{x\bar{x}}+
(\frac{\textbf{k}^2_{\perp}+m_c^2(x-\bar{x})^2}{2x\bar{x}})^2]e^{-\frac{\textbf{k}^2_{\perp}+
m_c^2(x-\bar{x})^2}{8x\bar{x}\alpha^2}}.
\end{eqnarray}
Applying the Fourier transform to replace the transverse momentum $\textbf{k}_{\perp}$ with its
conjugate variable $b$, the 3S oscillator wave function can be taken as
\begin{eqnarray}\label{eq:wave1}
\Psi_{(3S)}(x,\textbf{b})&\sim& \int \textbf{d}^2\textbf{k}_{\perp}e^{-i\textbf{k}_{\perp}
\cdot \textbf{b}}\Psi_{(2S)}(x,\textbf{k}_{\perp})\nonumber\\
&\propto &
x\bar{x}\mathcal {T}(x)
e^{-x\bar{x}\frac{m_c}{\omega}[\omega^2b^2+(\frac{x-\bar{x}}{2x\bar{x}})^2]},
\end{eqnarray}
with
\begin{eqnarray}
\mathcal {T}(x)=7-2\frac{m_c(x-\bar{x})^2}{\omega x\bar{x}}-24m_c \omega b^2x(1-x)+
(\frac{m_c(x-\bar{x})^2}{\omega x\bar{x}}-4b^2m_c\omega x\bar{x})^2.
\end{eqnarray}
We then propose the 3$S$ states distribution amplitudes
inferred from Eq.~(\ref{eq:wave1}),
\begin{eqnarray}
\Psi_{(3S)}(x,b)
\propto \Phi^{asy}(x)\mathcal {T}(x)
e^{-x\bar{x}\frac{m_c}{\omega}[\omega^2b^2+(\frac{x-\bar{x}}{2x\bar{x}})^2]},
\end{eqnarray}
with the $\Phi^{asy}(x)$ being the asymptotic models, which are given in \cite{plb612215}.
Therefore, we have the distribution amplitudes for the radially excited charmonium mesons
$\eta_c(3S)$ and $\psi(3S)$:
\begin{eqnarray}\label{eq:wave}
\Psi^{L,T,v}(x,b)&=&\frac{f^{(T)}_{3S}}{2\sqrt{2N_c}}N^{L} x\bar{x}\mathcal {T}(x)
e^{-x\bar{x}\frac{m_c}{\omega}[\omega^2b^2+(\frac{x-\bar{x}}{2x\bar{x}})^2]},\nonumber\\
\Psi^t(x,b)&=&\frac{f^{T}_{3S}}{2\sqrt{2N_c}}N^t (x-\bar{x})^2\mathcal {T}(x)
e^{-x\bar{x}\frac{m_c}{\omega}[\omega^2b^2+(\frac{x-\bar{x}}{2x\bar{x}})^2]},\nonumber\\
\Psi^V(x,b)&=&\frac{f_{3S}}{2\sqrt{2N_c}}N^V [1+(x-\bar{x})^2]\mathcal {T}(x)
e^{-x\bar{x}\frac{m_c}{\omega}[\omega^2b^2+(\frac{x-\bar{x}}{2x\bar{x}})^2]},\nonumber\\
\Psi^s(x,b)&=&\frac{f_{3S}}{2\sqrt{2N_c}}N^s \mathcal {T}(x)
e^{-x\bar{x}\frac{m_c}{\omega}[\omega^2b^2+(\frac{x-\bar{x}}{2x\bar{x}})^2]},
\end{eqnarray}
with the normalization conditions
\begin{eqnarray}
\int_0^1\Psi^{i}(x,0)d x&=&\frac{f^{(T)}_{3S}}{2\sqrt{2N_c}}\;.
\end{eqnarray}
$N_c$ above is the color number,  $N^{i}(i=L,t,V,s)$ are the normalization constants.
 $f_{3S}$ and  $f^T_{3S}$ are vector and tensor decay constants, respectively.
Since the energy spectrum of a
three-dimensional harmonic oscillator is given by $E_{nl}=[2(n-1)+l+\frac{3}{2}]\omega$,
the value of the frequency $\omega$ can be determined by  the difference between the two adjacent energy states.
Here, the parameter $\omega \approx (m_{4S}-m_{3S})/2\approx 0.1$ GeV.

\end{appendix}

%%\begin{appendix}
%%\end{appendix}
%%%%%%%%%%%%%%%%%%%%%%%%%%%%%%%%%%%%%%%%%%%%%%%%%%%%%%%%%%%%%%%%%%

\end{document}